# Theoretical simulation and experimental verification of dynamic caustic manipulation using a deformable mirror for laser material processing

Marco Smarra[1*], Evgeny L. Gurevich[2], Andreas Ostendorf[1]

[1]Ruhr-University Bochum, Faculty of Mechanical Engineering, Chair of Applied Laser Technologies, Universtiätsstr. 150, 44801 Bochum, Germany (*corresponding author: smarra@lat.rub.de)

[2]gurevich@fh-muenster.de, FH Münster, Laserzentrum, Stegerwaldstr. 39, 48565 Steinfurt, Germany

## Abstract
The influence of a deformable mirror on spatial light modulation in ultrafast lasers processing is demonstrated. The deformable mirror was integrated into an optical setup which contains an additional lens for generating a nearly linear focus shift in the focal plane behind the f-theta lens. The deformation of the mirror surface can be described by the Zernike terms Defocus, Astigmatism, and a combination of both, resulting in a cylindric lens behavior. The influence of the mirror surface deformation in this optical setup on the intensity distribution in the focal plane was simulated. From the simulation results, the caustic in the focal plane was calculated. The simulation results were compared to experiments using a picosecond laser with a maximum pulse energy of about 60 µJ. We demonstrate that the initial astigmatism of the raw beam can be reduced using the deformable mirror. High linearity in the focus shift ($R^2 > 98\ \%$) and the generation of elliptical/ line intensity distributions are shown. Line intensity distribution was used to demonstrate slit drilling application in thin metal foils.

## Introduction

Ultrafast lasers are used in a wide range of micromachining, like drilling [1–4], cutting [5], and ablation [6–10]. A central feature of the short interaction time between the laser pulse and the material is the low thermal influence which can be neglected in many applications [11]. However, high peak fluences deteriorate some ablation characteristics, especially the ablation efficiency [12,13]. Moreover, the ongoing development in higher pulse energies results in high peak power, increasing the optical elements' damage probability. Furthermore, with high peak fluences on the workpiece, the formation of X-Ray radiation increases, with all its adverse effects on the process and machine safety [14]. Beam shaping is one solution to optimize the intensity distribution and increase the ablation efficiency using high pulse energies [15].

Spatial beam shaping influences the caustic and, therefore, changes the intensity distribution within the beam [16] or transforms it into multiple beams [16,17]. In general, beam shaping can be done by two concepts: changing the amplitude and/or the phase of the incoming beam. In ultrashort pulse lasers, e.g., the change of the intensity distribution within the beam can lead to a Top-hat beam. Compared to a Gaussian intensity distribution, using a Top-hat beam decreases the pulse overlap in processing smooth edges of isolations channels [16]. Multiple beams in one dimension can be used for a cumulated ablation of grooves. In contrast, multiple beams in two dimensions can be used for parallel drilling of multiple holes simultaneously.

For spatial beam shaping, static and dynamic systems can be used. Examples of static methods are freeform lenses [18,19] or diffractive optical elements (DOE) [20]. Their advantage is the high transmission efficiency. In contrast, due to their static form, the flexibility for different ablation processes is not given.

Dynamic beam shaping methods allow the change of the beam shape, e.g., during the process or due to changes in the initial raw beam (e.g., heat effects), without changing the optical setup/components. There are several beam shaping methods; three of them will be discussed shortly.

First, e.g., a liquid inside a membrane changes its form due to pressure [21] or applied voltage [22]. These so-called liquid lenses can achieve large apertures. However, the horizontal orientation is crucial to suppress errors like astigmatism, which results from gravity forces to the liquid [23]. Second, Acousto-Optic-Modulators (AOM) offer the highest speed in beam shaping (up to several hundred kilohertz) [24]. Nevertheless, high-speed can only be achieved by small apertures [25]. Besides, AOM technology suffers from low damage thresholds [26]. Third, Spatial-Light-Modulators (SLMs) are used in a wide range of applications. They offer high flexibility and a medium speed (depending on the dimension of the SLM) [27]. However, damage and efficiency limitations occur in transmission, so it is more suitable to use reflective SLMs [28]. Due to their pixel structure, diffraction occurs and needs to be controlled or blocked [29].

In this paper, a deformable mirror (DM) is used [30]: This mirror is built by combining a thin high reflective, dielectric coated mirror substrate glued to a piezoelectric ceramic. One side of the piezoelectric ceramic is electrically segmented, allowing the control of the mirror deformation dynamically, see [30]. Its flexibility is lower compared to an SLM. The speed for changing the mirror surface can go up to 200 Hz – and is then limited due to its eigenfrequencies (shown for a larger aperture in [31]). Newer developments offer deforming speeds of about 2 kHz [32]. Like AOM, EOM, and most SLMs, the deformable mirror influences the phase of the incoming laser beam. The DM is described in the optical components section.

## Material and Methods (Experimental setup)
### Optical Components

To demonstrate the potentials of beam shaping by using a deformable mirror (DM), the DM was inserted onto a beamline between an ultrashort-pulse laser source (Trumpf TruMicro 5050, P=50 W, τ=8 ps, f=800 kHz, $E_{Pulse}$=62.5 µJ, λ=1030 nm) and the 2D-scanning system (ScanLab Hurryscan II, $f_\Theta$=100 mm). An additional lens ($f_L$) with a focal length of 450 mm at a distance of 610 mm behind the DM was used to provide a linear focus shift at the focal plane of the optical system; see Figure 1 and Optical Setup for details.

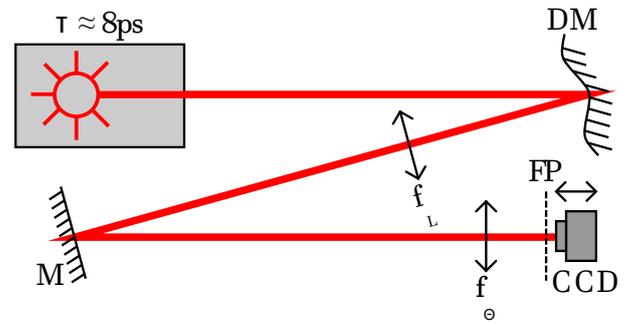

*Figure 1: Sketch of the optical setup. The phase of the picosecond laser beam is changed by the deformable mirror (DM). The additional lens ($f_L$) linearizes the focus shift in the focal plane (FP). The intensity distribution behind the scanning lens ($f_\Theta$) is measured using a CCD that moves along the beam propagation. For beam folding, flat mirrors (M) are used.*

### Deformable mirror (DM)

The DM used in this setup is a bimorph mirror [33], a piezoelectric ceramic combined with a thin mirror substrate. It contains 35 individual controllable segments. The continuous mirror surface provides a high reflectivity of more than 99.9 % without any diffraction. The deformation of the mirror surface is defined and controlled by Zernike-coefficients ($c$).

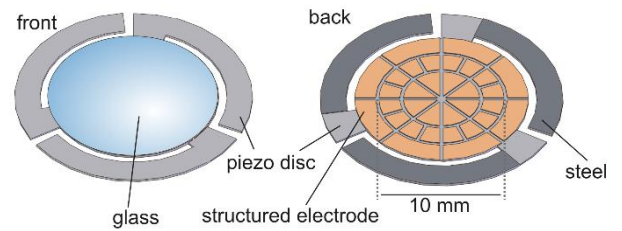

*Figure 2: Schematic of the deformable mirror used for the experiments: The AR-coated mirror substrate is glued to the front of a piezoelectric disc. The electrode on the backside of the piezoelectric disc is structured, allowing individual deformation control of each segment [33].*

In this study, the Defocus ($c_{Def}$) and astigmatism ($c_{Ast}$) aberrations of the deformable mirror surface on the beam in the focal plane are investigated. There are two Zernike coefficients for astigmatism $c_{Ast,0°}$ and $c_{Ast,45°}$. In this study, only Astigmatism 0° is analyzed so that the indices determining the angle will be dismissed if not necessary. Besides, a combination of both aberrations resulting in a Cylindric-like behavior of the deformable mirror surface is analyzed. The Cylindric-like deformation is no fundamental Zernike aberration, but as it is a combination of Defocus and Astigmatism in this study, a comparable notation will be used to represent the mirror deformation ($c_{Cyl}$). The Cylindric-like behavior will be achieved by

$$c_{Def} = 0.5 \cdot c_{Cyl} \qquad \text{Eq. 1a}$$

$$c_{Ast} = -c_{Cyl} \quad \text{Eq. 1b}$$

In Figure 3, the investigated deformations of the mirror surface are schematically shown. As the dimensions are not necessary for the fundamental understanding, it is essential to mention that the peak-to-valley values for all deformations are the same. For example, for the Astigmatism 0°, the mirror surface in the y-axis is bent down, while the mirror surface in the x-axis is bent up.

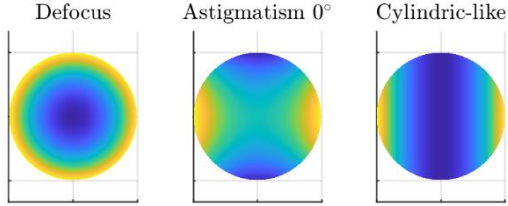

*Figure 3: Color-coded top view of three mirror deformations: Blue indicates low values, while yellow indicates high values. Although the dimensions are irrelevant for the basic understanding, the peak-to-valley values for all deformations are the same.*

### Theoretical Optical Simulation

In this paper, the influence of a deformed mirror surface on the beam propagation is analyzed. Therefore, in the beginning, two simulation methods are used. At first, the ABCD matrix is used for designing the optical setup. Therefore the deformable mirror's influence of the defocus deformation is replaced in the ABCD matrix by its lens behavior. The focal length of the replacing lens $(f_{DM})$ can be calculated by the diameter of the deforming mirror surface $(D)$ and the coefficient for the defocus $(c_{Def})$

$$f_{DM} = \frac{D^2}{32 \cdot c_{Def}} \quad \text{Eq. 2}$$

The Gaussian beam in the ABCD matrix is described by its q-parameter, which includes the radius of the wavefront $(R)$, the wavelength $(\lambda)$, the refractive index $(n)$, the beam radius $(w)$:

$$\frac{1}{q} = \frac{1}{R} - i\frac{\lambda}{\pi n w^2} \quad \text{Eq. 3}$$

The q-parameter changes due to the propagation in z from position *N* to *N+1* by

$$q_{N+1} = q_N + z \quad \text{Eq. 4}$$

The influence of an optical element at position *N* on the Gaussian beam can be calculated by

$$q_{N+1} = \frac{A \cdot q_N + B}{C \cdot q_N + D} \quad \text{Eq. 5}$$

For more complex mirror deformations, the diffraction integral was used. Here the influence of the mirror deformation on the phase and the aperture of optical components can be analyzed. The calculation of the diffraction integral was performed using the HPC (Campus Cluster) since each pixel of the output plane is influenced by every pixel from the input plane. For the optical components, the thin-element approximation (TEA) was used. TEA neglects the thickness of the optical component but uses only its resulting phase influence. The mirror deformation leads to a spatial influence on the phase, which is double the height of the deformation due to reflection.

### Experimental Optical Measurements

For the experimental results, the intensity distribution along the beam propagation near the nominal focus was measured. The measurement was done by a microscope objective, imaging the intensity distribution on a CCD camera. The output power of the laser was reduced to a minimum avoiding damage to the CCD camera. If necessary, additional neutral filters were used for further power reduction. By the intensity distribution image, the beam radius in the x- and y-plane can be determined. From these radii and the z-positions, the beam caustic can be determined.

### Optical Setup

For practical and laboratory reasons, the distance between the deformable mirror and the $f_\Theta$-lens was about 2 m. As a first step, the defocus deformation of the deformable mirror was used to determine the focus shift behind the $f_\Theta$-lens, see Figure 4. If the mirror is flat, the beam diameter remains nearly constant between the mirror and the $f_\Theta$-lens (green line). Assuming a convex mirror surface leads to an increasing beam diameter (red line). Respectively for a concave mirror surface, the beam reaches a waist before the lens (blue line). The vertical dashed line indicates the waist position. A mirror deformation in the range of $0 > c_{Def} > 3$ will produce a waist at the $f_\Theta$-lens, which may destroy the lens due to high intensity.

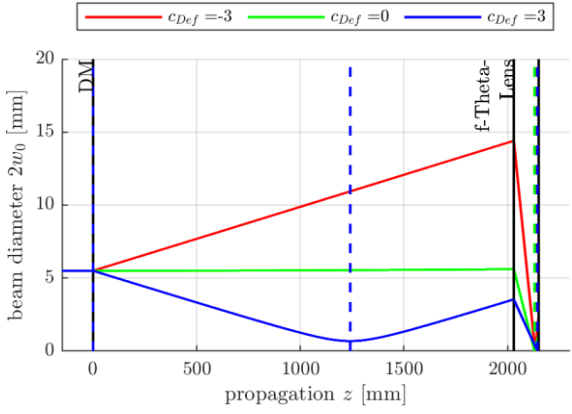

Figure 4: Optical Setup using a deformable mirror and an $f_\Theta$-lens results in a large variety of beam diameters at the $f_\Theta$-lens.

From Figure 4, the waist position behind the $f_\Theta$-lens can not be determined, so Figure 5 demonstrates the dependency of the waist position and the waist diameter behind the $f_\Theta$-lens on the mirror deformation. As expected for a flat mirror surface, the waist position equals the focal length of the $f_\Theta$-lens. However, it can be seen that neither the waist position nor the waist diameter shows monotonous behavior. Significantly the beam diameter increases due to a small illuminated area of the $f_\Theta$-lens, resulting in a larger effective focal length of the system. It would be suitable to achieve a linear focus shift behind the $f_\Theta$-lens. In addition, a linear behavior would be preferable for controlling the focus shift during material processing.

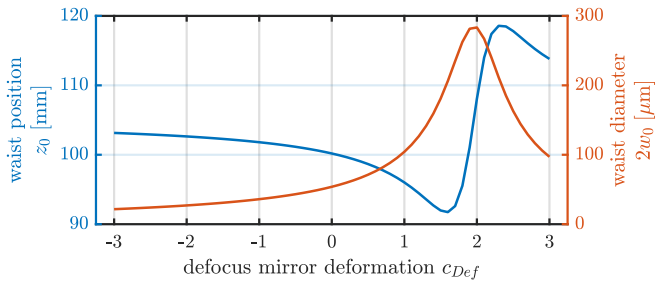

Figure 5: Detail simulation on waist position and waist diameter behind the $f\Theta$-lens depending on the mirror deformation.

An additional lens was added to the setup to provide a linear focus shift in the focal plane behind the $f_\Theta$-lens. The following specifications of the optical setup were made to choose the focal length and the position of the additional lens:

- When the defocus deformation of the mirror changes at $\Delta c_{Def} = 1$ the focus shift in the focal plane should be $\Delta z_0 \approx 1$ mm.
- The beam diameter at the $f_\Theta$-lens should be smaller than 11 mm and nearly independent of the mirror deformation to achieve a tiny spot in the focal plane.

A 4f-system or a telescope is not suitable to achieve these specifications: For the 4f-system and the telescope, the resulting focus shift in the focal plane is too high. Additionally, in the setup used for this study, the space for the second lens of a telescope was limited. These limitations lead to a setup using a single lens. The Gaussian-Matrix-Calculation (Eq. 5) was used to simulate the position and focal length of this lens. For the Matrix-Calculation, the Defocus deformation of the deformable mirror was replaced by a lens with a corresponding focal length. In this numerical simulation, the focal length of the additional lens and its position were varied. For each variation, the focus shift in the focal plane was calculated. Therefore, a change in the mirror deformation from $c_{Def} = 2.9$ to $c_{Def} = 3$ was estimated. The difference between these two waist positions is shown in Figure 6. The solid colored lines represent a constant focus shift. The dashed line represents the chosen parameters: A focal length of $f_L = 450$ mm in a distance of about $\Delta z_{DM,L} \approx 610$ mm.

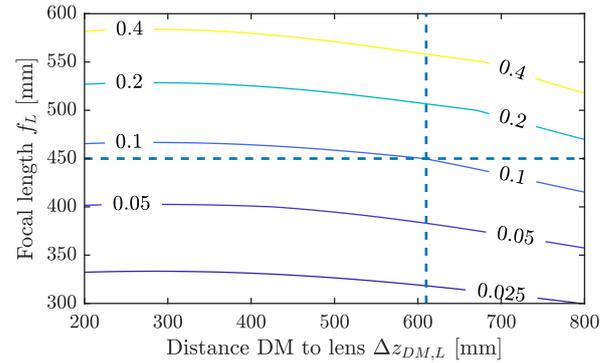

Figure 6: Focus shift (colored solid lines) in dependence on the focal length of the additional lens ($f_L$) and its distance to the deformable mirror ($\Delta z_{DM,L}$), when the defocus deformation changes from $c_{Def} = 2.9$ to $c_{Def} = 3.0$. Each solid line represents a constant focus shift in millimeters. The dashed lines represent the chosen parameters.

The beam propagation for chosen Defocus deformation of the deformable mirror is shown in Figure 7. Each beam propagation represents a different deformation of the deformable mirror surface. The vertical dashed lines represent the waist position for the corresponding beam. The additional lens leads to two main effects of the Defocus deformation of the deformable mirror at the $f_\Theta$-lens:

- The beam divergence at the $f_\Theta$-lens is changed, but
- the beam diameter on the $f_\Theta$-lens is nearly independent of the Defocus deformation of the deformable mirror.

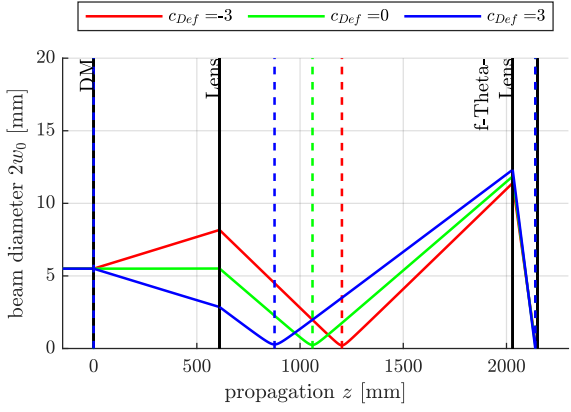

Figure 7: Beam propagation against the Defocus deformation of the mirror surface.

Figure 8 shows the dependence of the Defocus deformation of the deformable mirror surface on the waist position and the waist diameter in the focal plane. The dashed line shows the linear approximation. The additional lens changes the focal length of the optical system. The effective focal length increases from 100 mm of the f-Θ-lens to about 111 mm. The Defocus deformation of the mirror surface from $c_{Def} = -3$ to $c_{Def} = 3$ leads to a focus shift from $z_0 \approx 113.8$ mm to $z_0 \approx 109.4$ mm. For a change of the Defocus deformation of $\Delta c_{Def} = 6$, the waist position is shifted about $\Delta z_0 = 4.4$ mm. In this range, the linearity is high ($R^2 = 99.9\ \%$).

The waist diameter in the focal plane decreases over the Defocus deformation from $2w_0 \approx 30$ μm to $2w_0 \approx 27$ μm. This decrease is mainly based on the variation of the beam diameter and the divergence at the f-Θ-lens. The larger the diameter on the f-Θ-lens, the smaller the waist diameter. For this study, the range and linearity of the focus shift and the low variation of the waist diameter are suitable.

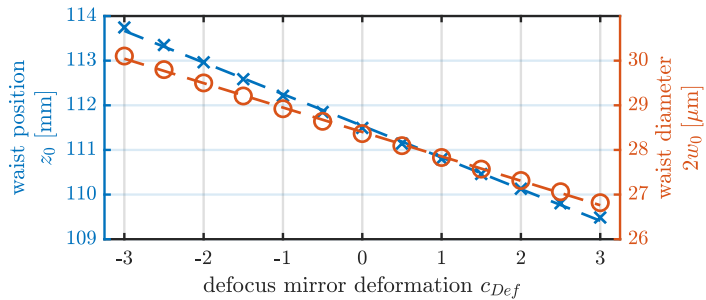

Figure 8: Waist position and waist diameter against the Defocus deformation of the deformable mirror surface. Dashed lines show the linear approximation.

### Simulation Setups

The optical setup was designed using the Gaussian-Matrix-Calculation (see Eq. 5) as this delivers fast results for an extensive range of parameter variations. The diffraction integral was calculated for the theoretical analysis of various deformations of the deformable mirror surface and their influence on the intensity distribution in the focal plane. The HPC was used due to the high number of calculations between the entrance plane (at the DM), additional lens, f-Θ-lens, and screen. The screen position can be considered as the workpiece surface. At the optical components, like additional lens and f-Θ-lens, the intensity distribution was calculated. Here, it is possible to analyze if there is any diffraction due to the size of the optical component.

The calculation of the diffraction integral offers the possibility to analyze more complex surface deformation such as astigmatism, trefoil, or coma. Besides, the results contain electromagnetic field information, which provides the opportunity to analyze interference patterns. Finally, by calculating the diffraction integral, implementing a closed-loop algorithm (e.g., Gerchberg-Saxton) to find the mirrors' surface deformation for a desired intensity distribution in the focal plane is possible.

For calculating the diffraction integral, TEA was used. In TEA, the optical elements, such as lenses, are described by their phase and amplitude influence on the incident beam. A lens, e.g., influences the phase of the incident beam by its focal length due to the thickness and refractive index. The amplitude of the incident beam is, e.g., influenced by the mirror mount (blocking) or the transparency of the coating (attenuation).

The intensity distribution near the focal plane was calculated at several positions before and behind the focal plane. Based on the intensity distribution in each z-position, the beam diameter ($2w$) was measured. With the beam diameter at different z-positions, the beam propagation can be calculated. From

$$w(z) = w_0 \cdot \sqrt{1 + \left(\frac{z - z_0}{z_R}\right)^2} \qquad \text{Eq. 6}$$

the waist position ($z_0$), waist radius ($w_0$), and Rayleigh length ($z_R$) can be determined. These parameters were analyzed in the x/z- and y/z-plane due to the astigmatic influence. The plane is represented in the indices of the parameter, such as the waist position (e.g. $z_{0,x}$).

### Experimental Analysis

The intensity distribution of the laser at different z-positions was measured using a microscope objective and a CCD camera. From the intensity distribution, the beam diameter was determined. Then, the beam propagation was recalculated by the beam diameters along the measured z-position. Finally, the waist position was determined from the recalculation.

## Results and Discussion

### Theoretical Simulation

*Defocus*

For each simulated intensity distribution along the z-propagation in the nearfield of the focal plane ($z = 0$ mm) the beam radius was determined. Figure 9 demonstrates the dependence of the Defocus deformation of the deformable mirror surface on the caustic (a: $c_{Def} = 2$) compared to the raw beam (b: $c_{Def} = 0$). The raw beam is calculated as a beam formed by a flat mirror surface of the deformable mirror. The beam radius in the x/z-plane is shown in black crosses, while the y/z-plane is shown in red triangles. By Eq. 6, the waist position (vertical line) and waist radius (horizontal line) are determined.

By the Defocus deformation of the deformable mirror surface $c_{Def} = 2$, the waists in the x/z- and y/z-planes are both shifted at the same distance in front of the focal plane. Changing the deformation of the deformable mirror surface can be compared with changing the mirrors' focal length and thus changing the optical systems' focal length. The flat mirror $(c_{Def} = 0)$ equals an infinitive focal length. The positive Zernike-coefficient leads to a mirror deformation which can be compared to a converging lens. So, a positive Zernike-coefficient result reduces the optical system's focal length. This reduction leads to a waist in front of the focal plane. Changing the direction of the deformation by negating the sign to negative Zernike-coefficients leads to a waist position behind the focal plane. Again, the waists in the x/z- and y/z-plane are at the same z-position behind the focal plane. The focal length of the optical system is longer for a negative Zernike-coefficient. Thus a negative Zernike-coefficient leads to behavior compared to a convex lens.

In Figure 9 a) and b), the caustic of two exemplary mirror deformations are shown. Figure 9 c) demonstrates the shift of the waist position depending on the defocus mirror deformation in the range between [-3; 3]. As expected by the Gaussian matrix simulation, the dependence of the focus shift on the defocus mirror deformation is nearly linear, shown by the linear approximation (dashed line, $R^2 = 99.1$ %). But, although the absolute value of the Zernike-coefficient is the same, the shift behind the focal plane $(c_{Def} = -2)$ is about $\Delta z_0 \approx 1.9$ mm, whereas the shift ahead of the focal plane $(c_{Def} = 2)$ is only about $\Delta z_0 \approx 1.6$ mm. In total, the waist can be shifted over a range of about $\Delta z_0 \approx 3.5$ mm by a change of the Defocus deformation of the deformable mirror surface, represented by the Zernike coefficient, of $\Delta c_{Def} = 4$. The z-position of the waists in the x/z- and y/z-plane are identical, resulting from the rotationally symmetric deformation of the deformable mirror surface.

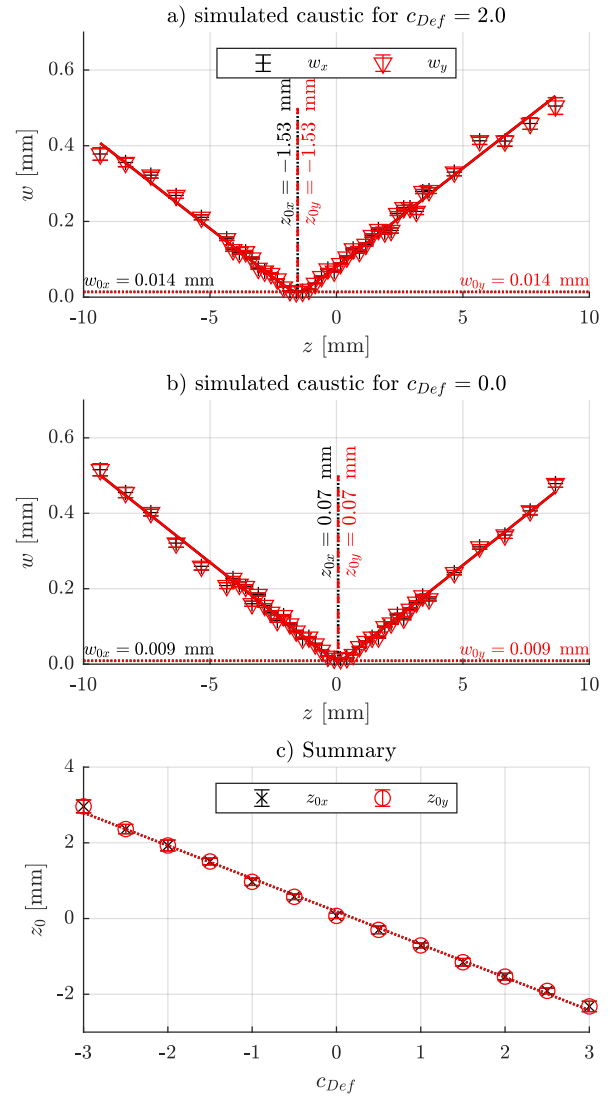

*Figure 9 a) and b): Simulated caustic for chosen defocus mirror surface deformation. c): Dependence of the waist positions on the defocus mirror surface deformation. Due to the rotation symmetry of the spot, the waist radius in the x/z- and y/z-plane are equal. Furthermore, the waist positions in the x/z- and y/z-plane are equal.*

*Astigmatism*

The same absolute values as for the Defocus are used for the astigmatic deformation of the deformable mirror surface. Figure 10 a) and b) demonstrate the dependence of the caustic in the nearfield of the focal plane on the astigmatic deformation of the deformable mirror surface. The raw beam can be compared to Figure 9 b). Again, the x/z- (black) and the y/z-plane (red) are shown.

For $c_{Ast} = 2$ the waist in the x/z-plane is shifted ahead of the focal plane. The waist in the y/z-plane is shifted behind the focal plane. A negative Zernike-coefficient switches the direction of the mirror surface deformation. Therefore, the direction of the waist shift in the x/z- and y/z-plane is switched, too. The amount of the shifting of the waists in the x/z- and y/z-plane is half the amount of the Defocus. As the value of the Zernike-coefficient describes the mirror surface deformation, the peak-to-valley value also depends on the deformation: For an astigmatic deformation of the deformable mirror surface, the deformation (e.g. $c_{Ast} = 2.0$) equals a converging lens in one plane (e.g., x/z) by bending up, while the perpendicular plane (in this example y/z) equals a convex lens by bending down. Therefore, in this example, a Zernike-coefficient for Astigmatism of $c_{Ast} = 2$ leads to a comparable Defocus of $c_{Def} = 1$ in the x/z-plane and $c_{Def} = -1$ in the y/z-plane. In total, this deformation leads to a difference in the Zernike-coefficient of 2. Thus, by an astigmatic mirror deformation, the waists in the x/z- and y/z-plane are shifted in opposite directions. The distance between the waist in the x/z- and y/z-plane is $\Delta z_{Ast} \approx 1.75$ mm, which is, again, half the amount of the Defocus shift. The beam radius in the focal plane increases by the same amount in the x/z- and y/z-plane. The increase results from the nearly linear increasing distances of the waists in the x/z- and y/z-planes the focal plane. The waist radius in each plane increases due to the astigmatic deformation of the deformable mirror surface.

Figure 10 c shows the focus shift of the x/z- (black) and y/z- (red) plane as well as the total distance between the two waists (magenta) in dependence on the astigmatic mirror deformation. The dashed lines indicate a linear approximation. Figure 10 c demonstrates the contrary shift of the waist in the x/z- and the y/z-plane with an increasing mirror deformation. Compared to the defocus mirror deformation, an increasing astigmatic mirror deformation leads to a linear waist shift in the x/z- and the y/z-plane ($R^2 = 99.8$ %). This linearity shows the analogy of the astigmatic deformation to the defocus deformation: One axis (e.g., x/z) of the mirror bends up, while the perpendicular axis (then y/z) bends down (see Figure 3). So, the mirror acts as a concave mirror in the x/z- and a convex mirror in the y/z-axis. Nevertheless, the total shift of the waist in the analyzed axis is half the shift of the defocus deformation.

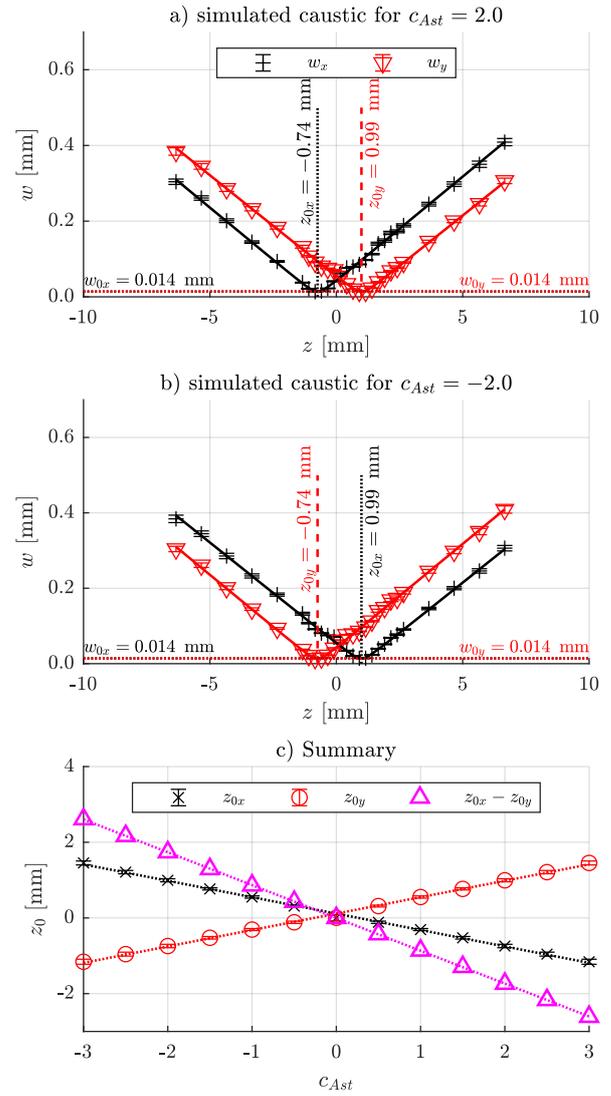

*Figure 10 a) and b): Simulated caustic for chosen astigmatic mirror surface deformation. c): Dependence of the waist positions on the astigmatic mirror surface deformation.*

*Cylindric-like behavior*
The Cylindric-like behavior combines the Defocus and astigmatic deformation of the deformable mirror surface (see Eq. 6 and Figure 3). As for the Defocus and Astigmatism, the same values for the deformation are used. Figure 11 a) and b) demonstrate the caustic's dependence in the nearfield of the focal plane on the Cylindric-like deformation of the deformable mirror surface. Again, the x/z- (black) and y/z-plane (red) are analyzed.

For the Cylindric-like behavior represented by $c_{Cyl} = 2$, the waist in the x/z-plane is shifted ahead of the focal plane. The position equals the focus shift resulting by $c_{Def} = 2$. For the x/z-plane, the behavior of the mirror can be compared to a convex lens. The waist in the y/z-plane remains in the focal plane. For the y/z-plane, the mirror behavior can be compared to a flat mirror.

Changing the direction of the deformation by negating the deformation coefficient($c_{Cyl} = -2$), leads to a focus shift of the waist in the x/z-plane behind the focal plane. Again, the position of the waist (in the x/z-plane) is the same as for the comparable Defocus ($c_{Def} = -2$). The waist in the y/z-plane remains in the focal plane, as seen before ($c_{Cyl} = 2$). Thus the behavior in the x/z-plane can now be compared to a concave lens, respectively, a flat mirror in the y/z-plane.

The beam radius in the y-direction of the focal plane is kept constant by the Cylindric-like deformation, while the beam radius in the x-direction of the focal plane increases. This increase is a consequence of the waist shift. So, the behavior of this deformation equals an elliptical or line shape in the focal plane. In the focal plane, the beam radius in the y-direction is much broader than the perpendicular x-direction. The distance between the waists modifies the aspect ratio of the elliptical shape.

Figure 11 c) shows the waist shift in the x/z- (black) and y/z-plane (red) as well as the total distance between the waist in dependence on the Cylindric-like mirror deformation. The dashed line demonstrates the linear approximation. As demonstrated in Figure 11 a) and b), the waist in the y/z-plane remains in the focal plane independent of the Cylindric-like mirror deformation. Moreover, in contrast to the y/z-plane, the waist in the x/z-plane is shifted linearly ($R^2 = 99.9\ \%$). This shift is similar to the waist shift using the defocus deformation. Furthermore, because the y/z-plane's waist is fixed, the total distance between the two waists equals the waist shift of the x/z-axis.

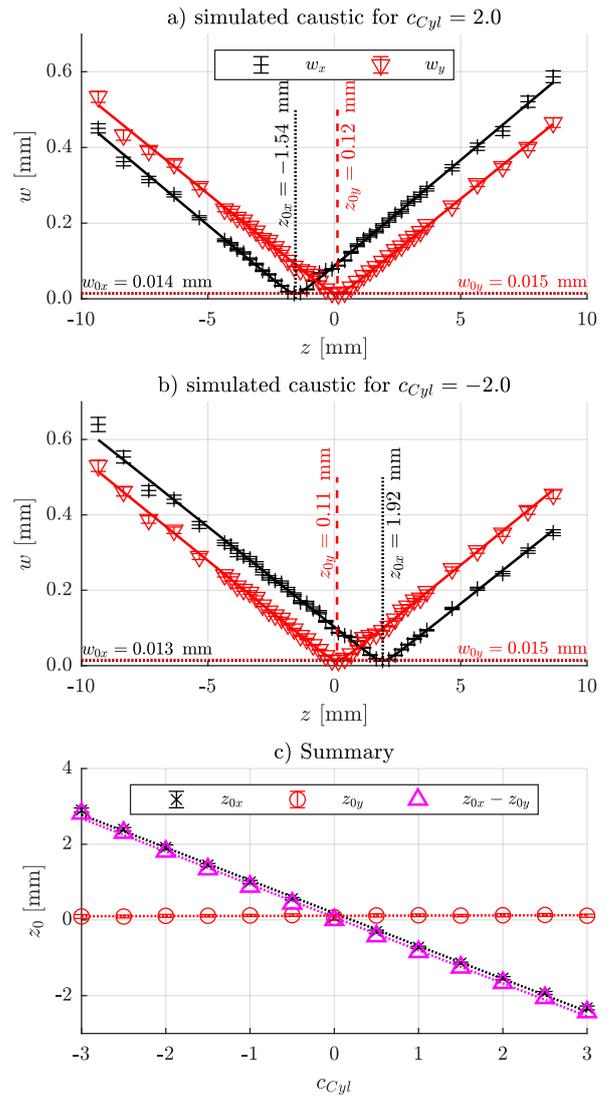

*Figure 11 a) and b): Simulated caustic for chosen Cylindric-like mirror surface deformation. c): Dependence of the waist positions on the Cylindric-like mirror surface deformation.*

### Overview of the simulated beam shapes generated by the mirror deformations

So far, the calculated caustics in the focal plane for defocus, astigmatism, and a combination of both are shown. From the caustic, the beam shape was estimated. Figure 12 shows the intensity distribution for selected astigmatic deformation and z-positions around the focal plane. The z-positions equal a Defocus deformation in the focal plane.

Beginning with the Defocus ($c_{Ast} = 0$), the beam radius increases if the distance between the waist and the focal plane increases. Due to the rotationally symmetric deformation of the deformable mirror surface, the beam radius increases uniformly in all directions of the focal plane, resulting in circular intensity distribution.

The astigmatic deformation ($z = 0$) of the deformable mirror surface leads to an increasing beam diameter in the focal plane. This increase was also estimated by Figure 10 due to the contrary shift of waists in the x/z- and y/z-plane. Nevertheless, the intensity distribution is not circular compared to the waist shift occurring by the Defocus deformation of the deformable mirror. The intensity distribution is not associated with an effect due to the numerical simulation. Instead, it is a result of the interference of the convergent and divergent parts of the beam.

The combination of the Defocus and astigmatic deformation described in Eq. 6 leads to a Cylindric-like behavior. The waist in one plane (e.g., x/z) is shifted and in the perpendicular plane (thus y/z) remains in the focal plane, as shown in Figure 11. The result is an elliptical or line shape with a high aspect ratio ($z = 1$ mm and $c_{Ast} = 2$) and ($z = -1$ mm and $c_{Ast} = -2$).

As additional information, the rotation of the elliptical orientation can be achieved by rotating the astigmatic part of the mirror deformation. For the rotation of astigmatism around the z-axis by the angle $\beta$, a combination of the Zernike-coefficients for Astigmatism 0° and 45° needs to be combined by

$$c_{Ast,0°} = c_{Ast,\beta} \cdot \cos 2\beta \qquad \text{Eq. 7a}$$

$$c_{Ast,45°} = c_{Ast,\beta} \cdot \sin 2\beta \qquad \text{Eq. 7b}$$

Thus, the rotation of astigmatism around the z-axis with the angle $\beta = 90°$ leads to a negative Zernike-coefficient for Astigmatism 0° ($c_{Ast,0°}$) while the Zernike-coefficient for Astigmatism 45° is zero. For the rotation of the Cylindric-like behavior, Eq. 1b needs to be updated. The rotation of the Cylindric-like behavior by $\beta = 90°$ leads to the elliptical or line intensity distribution as shown in Figure 12 ($z = -1$ mm and $c_{Ast} = 2$) and ($z = 1$ mm and $c_{Ast} = -2$).

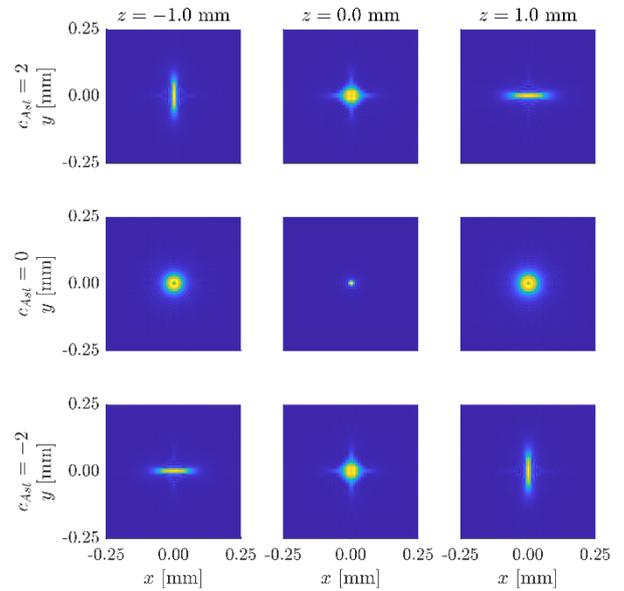

Figure 12: Intensity distribution in the focal plane as a function of the defocus and astigmatic deformation of the deformable mirror surface.

### Experimental Results

For the simulation results, the influence of the mirror deformation on the propagation was shown. This section shows the experimental results depending on the mirror surface deformation. Therefore, the intensity distribution was captured by a CCD camera and microscope-objective at different positions in the nearfield of the focal plane. From each image, the beam diameter was measured, and the propagation along the z-axis was calculated. From the fitting of the gaussian propagation, see Eq. 6, the waist positions were determined. The determined waist positions are shown in dependence on the mirror surface deformation. Dashed lines indicate a linear approximation.

In contrast to the simulation, the range of the mirror deformation was limited to $c_{Def} = [-2; 2]$. A further increase of the mirror surface deformation – especially for the astigmatic mirror surface deformation - led to low intensities on the CCD camera and, therefore, to errors in the determination of the beam caustic.

#### Raw beam analysis and correction

The simulation results were performed with an idealized laser beam (e.g., M²=1). The experiments were done using a laser beam, showing slight astigmatism (see Figure 13 a). The distance between the perpendicular waists is about $\Delta z\_Ast \approx 250$ µm. The waist distance was reduced using an astigmatic mirror deformation (see Figure 13 b). With this mirror deformation, the waist radius of the corrected raw beam increases in each plane which can be seen best for the y/z-plane. This astigmatism compensating mirror deformation was referred to as the new initial mirror surface. The following Zernike values are added to this initial deformation.

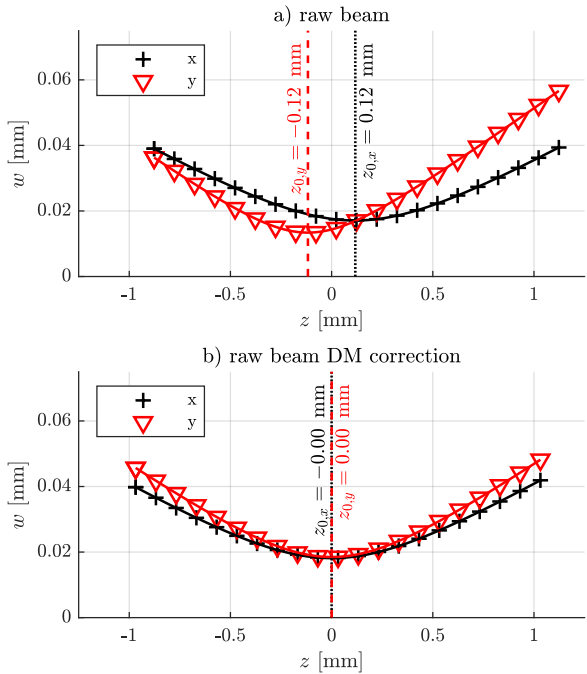

Figure 13: Raw beam compare: a) with flat mirror and b) with astigmatic correction using the DM.

Note that all simulations were done under idealized circumstances, especially perpendicular angle of incident on the deformable mirror and no astigmatism on the incident laser beam. A perpendicular incident angle on the deformable mirror can be achieved for linear polarized lasers, using half and quarter waveplates combined with a polarizing beam splitter. However, this setup cannot be used for random polarized lasers. For the experimental setup, a small angle of incident of fewer than 3 degrees was used. With the correction of initial astigmatism, the small incident angle does not significantly disturb the laser beam.

*Defocus*

In Figure 14, the waist position depending on the Defocus deformation of the deformable mirror is shown. The Zernike-coefficient represents the deformation of the mirror surface for Defocus ($c_{Def}$). The x/z-(black cross) and y/z-plane (red circle) are analyzed separately due to the comparability to the following results. The simulation results showed the same waist position in the x/z- and y/z-plane due to the rotational symmetry of the defocus deformation.

The waist position can be shifted nearly linearly by the Defocus deformation of the deformable mirror surface, indicated by the dashed line, representing a linear approximation with $R^2 = 98.7\ \%$. The simulation shows that the shift is a little more pronounced for positive Zernike-coefficients than negative Zernike-coefficients. For further investigation, the deviation from a linear slope is acceptable. As seen in the simulation, the determined shift of the waist in the x/z- and the y/z-plane is the same. Thus, the experimental results fit the simulation results (see Figure 9 c).

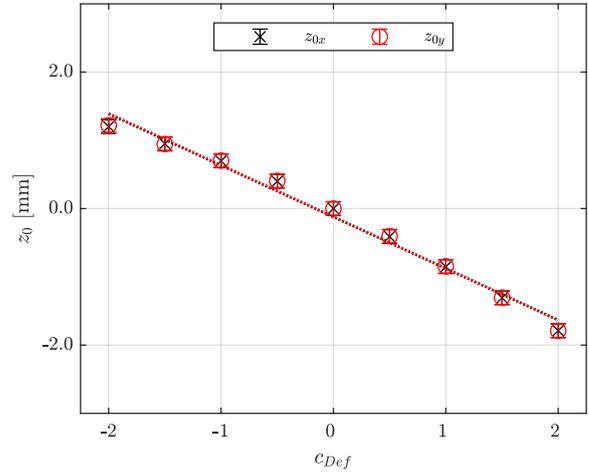

Figure 14: Experimental results of the defocus mirror-surface deformation on the waist position in the nominal focus plane.

*Astigmatism*

Figure 15 demonstrates the waist position depending on the astigmatic mirror surface deformation. Here again, the x/z- and y/z-plane are shown (black crosses and red circles). In addition, the total distance between the two waists is demonstrated in magenta triangles, as shown in Figure 10 c).

An increasing Zernike-coefficient shifts the waist in the y/z-plane in the positive direction. The waist in the x/z-plane is shifted in the negative direction. The increase is as linear as it is shown for the Defocus. The total z-distance between the waist in the x/z- and y/z-plane $(\Delta z_{Ast} = z_{0x} - z_{0y})$ increases linear $(R^2 = 99.9\ \%)$ as well with nearly the same slope as the Defocus. Again, the experimental results are in correspondence with the simulation results (see Figure 10 c).

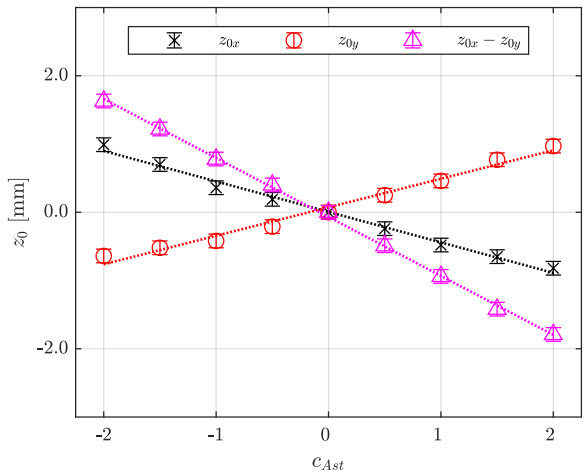

*Figure 15: Experimental results of the astigmatic mirror-surface deformation on the waist position in the nominal focus plane.*

### Cylindric-like deformation

In correspondence to the experimental results for the astigmatic mirror surface deformation, Figure 16 summarizes the dependence of the Cylindric-like mirror surface deformation on the waist positions in the x/z- (black crosses) and y/z-plane (red circles). Again the total distance between the two waists is shown in magenta triangles.

As shown in the simulation results, the waist in the y/z-plane remains in the focal plane. Furthermore, the waist in the x/z-plane is shifted in dependence on the Cylindric deformation coefficient. This shift is nearly linear as expected from the simulation and shown in the defocus and astigmatic experimental results ($R^2 = 99.5\%$). Because the waist in the y/z-plane remains in the focal plane, while the waist in the perpendicular plane is shifted linearly, the total distance between the two perpendicular waists equals the waist position in the x/z-plane. The experimental results fit the simulation results.

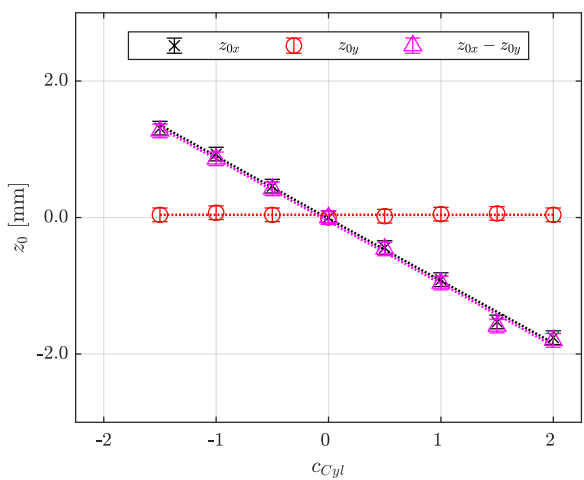

*Figure 16: Experimental results of the Cylindric-like mirror-surface deformation on the waist position in the nominal focus plane.*

### Application example

For demonstrating possible applications and the flexibility of the deformable mirror, elliptical holes are drilled in 20 µm thick stainless steel samples. Figure 17 shows a backlight image of the processed sample. The shapes of the drilled holes follow the intensity distribution of the laser beam on the workpiece surface. The intensity distribution was changed for each row. A high aspect ratio of more than 1:10 for the hole geometry was achieved [4].

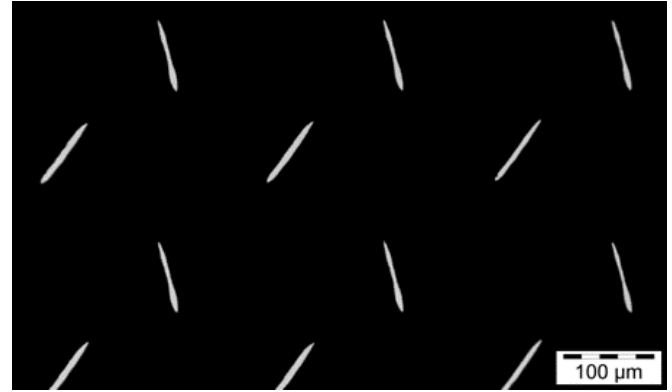

*Figure 17: Backlight image of elliptical holes drilled in 20 µm thick stainless steel. The shape of the holes follows the intensity distribution on the workpiece sample. The deformable mirror changed the intensity distribution in each row [4].*

## Conclusion and Outlook

In this study, we have shown the potentials in spatial beam shaping by using a deformable mirror. First, the simulation results calculating the diffraction integral show the possible influences of the mirror deformation on the beam propagation in the focal plane and its intensity distribution. Second, the simulation results were compared to the experimental study of a picosecond laser. It was shown that both – simulation and experiment – show the same behavior. Finally, it is demonstrated that a nearly linear focus shift is achieved due to the optical setup, $R^2 > 98\%$. For dynamic control of the focal position during material processing, this linear dependence is essential. The astigmatic and Cylindric-like mirror deformation also shows this linearity. Nevertheless, in contrast, these mirror deformations manipulate the z-distance of the waists in the x/z and y/z planes in opposite directions.

The influences of the deformable mirror surface can be adapted to a static mirror: high pressure by mounting optics also leads to a static deformation of the mirror surface and, therefore, influences the beam profile. Cylindric-like behavior can also be achieved by a not perpendicular angle when trespassing a lens or concave/convex mirror.

## Acknowledgment

The authors would like to thank the Photonics Laboratory of the University of Applied Sciences for providing the deformable mirror. Additional thanks go to Markus Gilbert from the HPC Campus-Cluster of the University of Applied Sciences Münster for helping to solve the diffraction integral using "octave" on multiple cores to reduce calculation time.

## Funding

This work was part of the project "Development of modern techniques for surface structuring with applications in regional industry" (MOVERO), funded by the INTERREG project of the European Union (INTERREG V A Programm 142091 MOVERO).

## Disclosure

The authors declare no conflicts of interest.

## References

[1] Abt F, Eichenberger M, Förster D. Laser drilling of banknote substrates. Results in Optics 2021;3:100058. https://doi.org/10.1016/j.rio.2021.100058.

[2] Watanabe W, Li Y, Itoh K. [INVITED] Ultrafast laser micro-processing of transparent material. Optics & Laser Technology 2016;78:52–61. https://doi.org/10.1016/j.optlastec.2015.09.023.

[3] Wlodarczyk KL, Brunton A, Rumsby P, Hand DP. Picosecond laser cutting and drilling of thin flex glass. Optics and Lasers in Engineering 2016;78:64–74. https://doi.org/10.1016/j.optlaseng.2015.10.001.

[4] Smarra M, Strube A, Dickmann K. Micro drilling using deformable mirror for beam shaping of ultra-short laser pulses. In: Klotzbach U, Washio K, Arnold CB, editors. SPIE LASE. SPIE; 2016, p. 97360.

[5] He C, Zibner F, Fornaroli C, Ryll J, Holtkamp J, Gillner A. High-precision Helical Cutting Using Ultra-short Laser Pulses. Physics Procedia 2014;56:1066–72. https://doi.org/10.1016/j.phpro.2014.08.019.

[6] Karnakis D, Rutterford G, Knowles M, Dobrev T, Petkov P, Dimov S. High quality laser milling of ceramics, dielectrics and metals using nanosecond and picosecond lasers. In: Okada T, Arnold CB, Meunier M, Holmes AS, Geohegan DB, Träger F et al., editors. Lasers and Applications in Science and Engineering. SPIE; 2006, p. 610604.

[7] Engelhardt U, Hildenhagen J, Dickmann K. Micromachining using high-power picosecond lasers. LTJ 2011;8(5):32–5. https://doi.org/10.1002/latj.201190056.

[8] Bruening S, Hennig G, Eifel S, Gillner A. Ultrafast Scan Techniques for 3D-µm Structuring of Metal Surfaces with high repetitive ps-laser pulses. Physics Procedia 2011;12:105–15. https://doi.org/10.1016/j.phpro.2011.03.112.

[9] Smarra M, Janitzki M, Dickmann K. Beam Shaping in Ultra-short Pulse Laser Processing for Enhancing the Ablation Efficiency. Physics Procedia 2016;83:1145–52. https://doi.org/10.1016/j.phpro.2016.08.120.

[10] Hildenhagen J, Engelhardt U, Smarra M, Dickmann K. Material specific effects and limitations during ps-laser generation of micro structures. In: Heisterkamp A, Meunier M, Nolte S, editors. SPIE LASE. SPIE; 2012, p. 824711.

[11] Nolte S, Momma C, Jacobs H, Tünnermann A, Chichkov BN, Wellegehausen B et al. Ablation of metals by ultrashort laser pulses. J. Opt. Soc. Am. B 1997;14(10):2716. https://doi.org/10.1364/JOSAB.14.002716.

[12] Raciukaitis G. Use of High Repetition Rate and High Power Lasers in Microfabrication: How to Keep the Efficiency High? JLMN 2009;4(3):186–91. https://doi.org/10.2961/jlmn.2009.03.0008.

[13] Neuenschwander B, Bucher GF, Nussbaum C, Joss B, Muralt M, Hunziker UW et al. Processing of metals and dielectric materials with ps-laserpulses: results, strategies, limitations and needs. In: Niino H, Meunier M, Gu B, Hennig G, editors. Laser Applications in Microelectronic and Optoelectronic Manufacturing XV. SPIE; 2010, 75840R.

[14] Legall H, Bonse J, Krüger J. Review of X-ray exposure and safety issues arising from ultra-short pulse laser material processing. J Radiol Prot 2020. https://doi.org/10.1088/1361-6498/abcb16.

[15] Smarra M, Dickmann K. Enhancing ablation efficiency in micro structuring using a deformable mirror for beam shaping of ultra-short laser pulses. In: Klotzbach U, Washio K, Arnold CB, editors. SPIE LASE. SPIE; 2016, p. 97360.

[16] Rung S. Laserscribing of Thin Films Using Top-Hat Laser Beam Profiles. JLMN 2013;8(3):309–14. https://doi.org/10.2961/jlmn.2013.03.0021.

[17] Büsing L, Eifel S, Loosen P. Design, alignment and applications of optical systems for parallel

processing with ultra-short laser pulses:91310. https://doi.org/10.1117/12.2051614.
- [18] Hoffnagle JA. Beam shaping with a plano-aspheric lens pair. Opt. Eng 2003;42(11):3090. https://doi.org/10.1117/1.1613957.
- [19] Laskin A, Šiaulys N, Šlekys G, Laskin V. Beam shaping imaging system for laser microprocessing with scanning optics. In: Reutzel EW, editor. Laser Material Processing for Solar Energy Devices II. SPIE; 2013, 88260F.
- [20] Barthels T, Reininghaus M. High precision ultrashort pulsed laser drilling of thin metal foils by means of multibeam processing. In: Dudley A, Laskin AV, editors. Laser Beam Shaping XVIII. SPIE; 2018, p. 10.
- [21] Ren H, Fox D, Anderson PA, Wu B, Wu S-T. Tunable-focus liquid lens controlled using a servo motor. Opt. Express 2006;14(18):8031. https://doi.org/10.1364/OE.14.008031.
- [22] Berge B, Peseux J. Variable focal lens controlled by an external voltage: An application of electrowetting. The European Physical Journal E 2000;3(2):159–63. https://doi.org/10.1007/s101890070029.
- [23] Blum M, Büeler M, Grätzel C, Aschwanden M. Compact optical design solutions using focus tunable lenses. In: Optical Design and Engineering IV. SPIE; 2011, 81670W.
- [24] Bechtold P, Hohenstein R, Schmidt M. Beam shaping and high-speed, cylinder-lens-free beam guiding using acousto-optical deflectors without additional compensation optics. Opt. Express 2013;21(12):14627. https://doi.org/10.1364/OE.21.014627.
- [25] Heberle J, Bechtold P, Strauß J, Schmidt M. Electro-optic and acousto-optic laser beam scanners. In: Klotzbach U, Washio K, Arnold CB, editors. SPIE LASE. SPIE; 2016, p. 97360.
- [26] Klotzbach U, Washio K, Arnold CB (eds.). SPIE LASE. SPIE; 2016.
- [27] Beck RJ, Waddie AJ, Parry JP, Shephard JD, Taghizadeh MR, Hand DP. Adaptive Laser Beam Shaping for Laser Marking using Spatial light Modulator and Modified Iterative Fourier Transform Algorithm. Physics Procedia 2011;12:465–9. https://doi.org/10.1016/j.phpro.2011.03.158.
- [28] Weiner AM. Femtosecond pulse shaping using spatial light modulators. Review of Scientific Instruments 2000;71(5):1929–60. https://doi.org/10.1063/1.1150614.
- [29] Kuang Z, Perrie W, Liu D, Edwardson S, Cheng J, Dearden G et al. Diffractive multi-beam surface micro-processing using 10ps laser pulses. Applied Surface Science 2009;255(22):9040–4. https://doi.org/10.1016/j.apsusc.2009.06.089.
- [30] Verpoort S, Wittrock U. Actuator patterns for unimorph and bimorph deformable mirrors. Appl. Opt. 2010;49(31):G37. https://doi.org/10.1364/AO.49.000G37.
- [31] Rausch P, Verpoort S, Wittrock U. Unimorph deformable mirror for space telescopes: environmental testing. Opt Express 2016;24(2):1528–42. https://doi.org/10.1364/OE.24.001528.
- [32] Verpoort S, Bittner M, Wittrock U. Fast focus-shifter based on a unimorph deformable mirror. Appl Opt 2020;59(23):6959–65. https://doi.org/10.1364/AO.397495.
- [33] Verpoort S, Rausch P, Wittrock U. Characterization of a miniaturized unimorph deformable mirror for high power CW-solid state lasers. In: Olivier SS, Bifano TG, Kubby J, editors. MEMS Adaptive Optics VI. SPIE; 2012, p. 825309.